\documentclass[hyperref,urlcolor=black]{article}
\usepackage{makeidx}
\usepackage[colorlinks=true,citecolor=black]{hyperref}
\usepackage{hyperref}
\usepackage{multirow}
\usepackage{multicol}
\usepackage[dvipsnames,svgnames,table]{xcolor}
\usepackage{graphicx}
\usepackage{epstopdf}
\usepackage{ulem}
\usepackage{hyperref}
\usepackage{amsmath}
\usepackage{amssymb}
\usepackage{enumerate}
\usepackage{latexsym}
\usepackage[a4paper]{geometry}
\usepackage{caption}
\usepackage{subfigure}
\usepackage{url}
\renewcommand{\thesubfigure}{\arabic{subfigure}} 
\makeatletter
\renewcommand{\@thesubfigure}{(\thesubfigure)\space}
\renewcommand{\p@subfigure}{} 
\makeatother
\hypersetup{
colorlinks=true,
linkcolor=black
}
\begin{document}
\captionsetup[figure]{labelfont={bf},name={Fig.},labelsep=period}
\title{\bf Some Issues on the Theory of the Mimic-Computing-Oriented Automata}

\date{}
\author{\large ZHU Wei-Jun\\
{\sffamily\small School of Information Engineering, Zhengzhou University, Zhengzhou  450001, China}
}
\maketitle
{\noindent\small{\bf Abstract:}
    A mimic computing oriented automaton can directly portray the behaviors of a mimic computing system. In this paper, we investigate the following theoretical problems on this type of automata: operational semantics and computational ability. First, we model a systematic structure of mimic computing via a mimic automaton. Second, we propose the operational semantics of the automaton in this scene. Third, the computational ability of this type of automata is studied.}

\vspace{1ex}

\section{Introduction}
China has developed the world’s first prototype of mimic computers under the leadership of Academician Wu Jiangxing in September 2013 \cite{W-MC}. ``The mimic computing aims to obtain the essence of the high-efficiency of computing through the multidimensional-reconstruction-based functional i.e., mimic variety, architecture.'' \cite{W-MC-J}. It aims to improve the efficiency of running by applying the idea that the application determines the structure and the structure determines the performance \cite{W-MC-J}. The results from a third-party's test show that the ratio of the efficiency of mimic computing raises 13.6-315 times on more than 500 scenarios of web service, N-body and image recognition \cite{W-MC}. Therefore, the mimic computer was selected into China's top ten scientific and technological advances for 2013 by the Chinese academicians \cite{top-10}.

Some important applications, such as the mimic web server \cite{Web} and the mimic mobile communication \cite{MSISDN}, have been proposed and constructed due to many advantages of the mimic computing.

Automaton theory has been proposed for some decades. Automata provide a mathematical model which portrays some systemic behaviors, for the different computing systems which have the different expressive abilities. Furthermore, the existing studies on automata also provide some theoretical basis and formal tools for the development of compilers. In recent years, various automata have been applied to model checking and formal verification. And these automata have been widely used in a series of fields, such as CPU design, network protocol verification, security protocol verification and software engineering analysis.

To the best of our knowledge, few studies have been done on the theory of the mimic-computing-oriented automata. This paper is aiming to address this problem.
\section{Mimic Automata}
We have proposed the mimic automata to establish a formal model for the mimic computing in our previous study \cite{MA}.

In short, a mimic automaton (MA) is composed of sequential automata (SA), cellular automata (CA) \cite{CA}\cite{SA} and hierarchical automata (HA) \cite{HA}\cite{HA1} according to certain logical relationship. In this paper, a SA can be a Turing machine (TM), a linear bounded automaton (LBA), a push-down automaton (PA) or a finite state automaton (FSA). What is the SA in a specific application? It is decided by the requirement of this application. The word ``Sequential" means that a formal machine needs to go through state transitions step-by-step.

In a mimic automaton, the SA describes the transitions of the systematic state, while the CA describes the dynamic changes in the structure of the execution bodies participating computing. The HA in MA only describes the hierarchical compositional relationship among different SA or CA. And it does not involve the internal states and state transitions of the SA. In fact, we take a HA as the macro-skeleton of a mimic automaton. Furthermore, the SA and CA are pieces of muscle attached to the skeleton. A formal model based on MA is obtained in this way.

In other words, a sequential automaton is responsible for the specific computing task, while a cellular automaton expresses dynamic heterogeneity, i.e., dynamic reconstruction of execution bodies. In addition, a hierarchical automaton expresses granularity, i.e., different granularity of execution bodies and different levels of state transitions. If a cellular automaton is replaced by a random cellular automaton, it can describe the stochastic dynamic reconstruction of execution bodies. Therefore, a mimic automaton formalizes some state transitions in dynamic, heterogeneous and random architectures at different granularity. The combination of sequential automata, cellular automata, random cellular automata and hierarchical automata makes a mimic automaton show the characteristics of the mimic variants. It should be noted that the essential properties and characteristics of mimic computing are dynamic, heterogeneous, random and the mimic variants.
\section{A modeling scene of DHR structures: the MA operational semantics}
In practical applications, the key of the mimic computing is dynamic heterogeneous. And it does not limit a specific architecture model. For the readability, we study the operational semantics of a mimic automaton with a specific application scenario.

A mimic automaton is used to model a typical mimic computing system. On the one hand, the dynamic and heterogeneous mechanism is the core ideas of mimic computing and mimic defense. On the other hand, there are significant differences between the principles of the mimic computing and that of the mimic defense, according to \cite{DHR} and \cite{CyberMS}. It should be noted that the DHR structure in this paper and the one in \cite{MA} are related to the mimic computing rather than the mimic defense. And Ref. \cite{MA} gives a brief description for modeling the scene of DHR structures. In order to study the operational semantics, we describe how to model this scenario step by step. 
\begin{figure}
	\includegraphics[width= \textwidth]{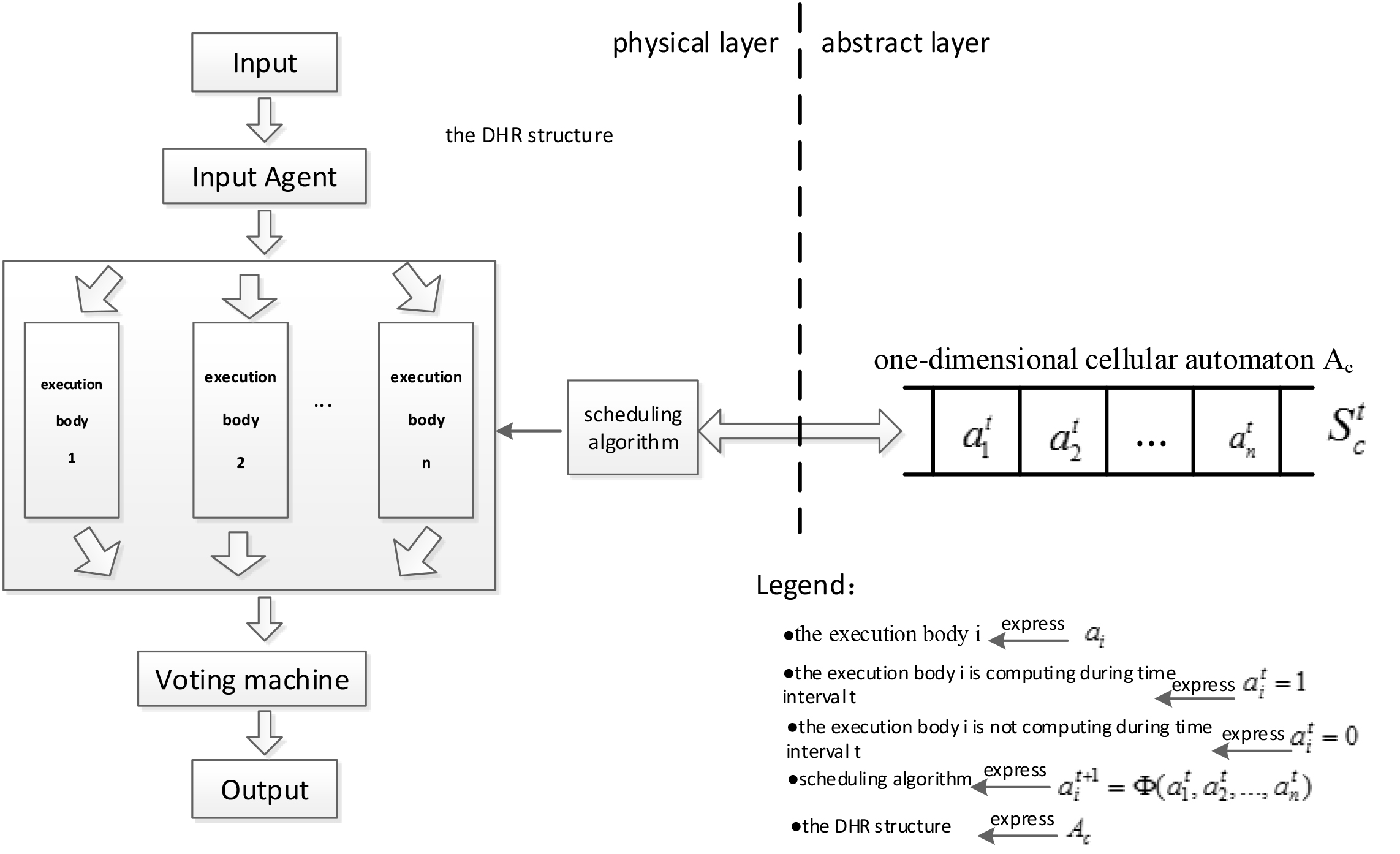}
	\caption{An one-dimensional cellular automaton for modeling a DHR structure}
	\label{fig1}
\end{figure}

A simple DHR structure is illustrated in the left side of Fig. \ref{fig1}. This DHR structure can be modeled by using an one-dimensional cellular automata which is a component in a mimic automaton, as shown in the right side of Fig. \ref{fig1}.
 
{\bf Example 1}
 
An one-dimensional cellular automaton $A_C$ is employed to characterize the dynamic changes in the structure of execution bodies, where an execution body is a hardware unit for computing. 

Supposing that the scheduling algorithm indicates that the $2^{th}$ execution body instead of other $n-1$ execution bodies computes at the present time, i.e., the time point t, the state of $A_C$ is $(0,1,0,..., 0)$ at the time point t. And we have $\chi^t=2$ at this time. Supposing that the scheduling algorithm indicates that the execution body $n^{th}$ instead of other $n-1$ execution bodies computes at the next time, i.e., the time point t+1, the state of $A_C$ is $(0,0,0,..., 1)$ at the time point t+1. And we have $\chi^{t+1}=n$ at that time.

In this example, the state of $A_C$ at the time point t, denoted by $S_C^t$, points out the execution body which is computing at that time. And the function $\Phi$ of $A_C$ is the mathematical model of the scheduling algorithm.

An one-dimensional cellular automaton $A_C$ only characterizing the following two points.
\begin{enumerate}[(1)]
\item Which hardware units, i.e., execution bodies, performs some computing tasks at the current time;
\item Which execution bodies take part in or exit the computing, as time goes by. 
\end{enumerate}

As for the specific computing process, $A_C$ does not involve how to compute on a given execution body, while it does not involve the systematic states and state transitions too. 

In Fig. \ref{fig1}, the scheduling algorithm schedules a number of execution bodies, which perform some computing tasks, in the dynamical and reconstructed way. The relationship between the DHR structure and a cellular automaton is as follows. A cell unit of the automaton represents an execution body. And the transition function $\Phi$ of the automaton represents the scheduling algorithm.  $\Box$

In a word, a cellular automaton is used to describe a simple dynamic DHR structure. On the basis of it, one can use a mimic automaton to formally describe the mimic computing. 

In a more complex DHR structure, different DHRs may be organized in a logical relationship.      

{\bf Example 2}
\begin{figure}
	\includegraphics[width= \textwidth]{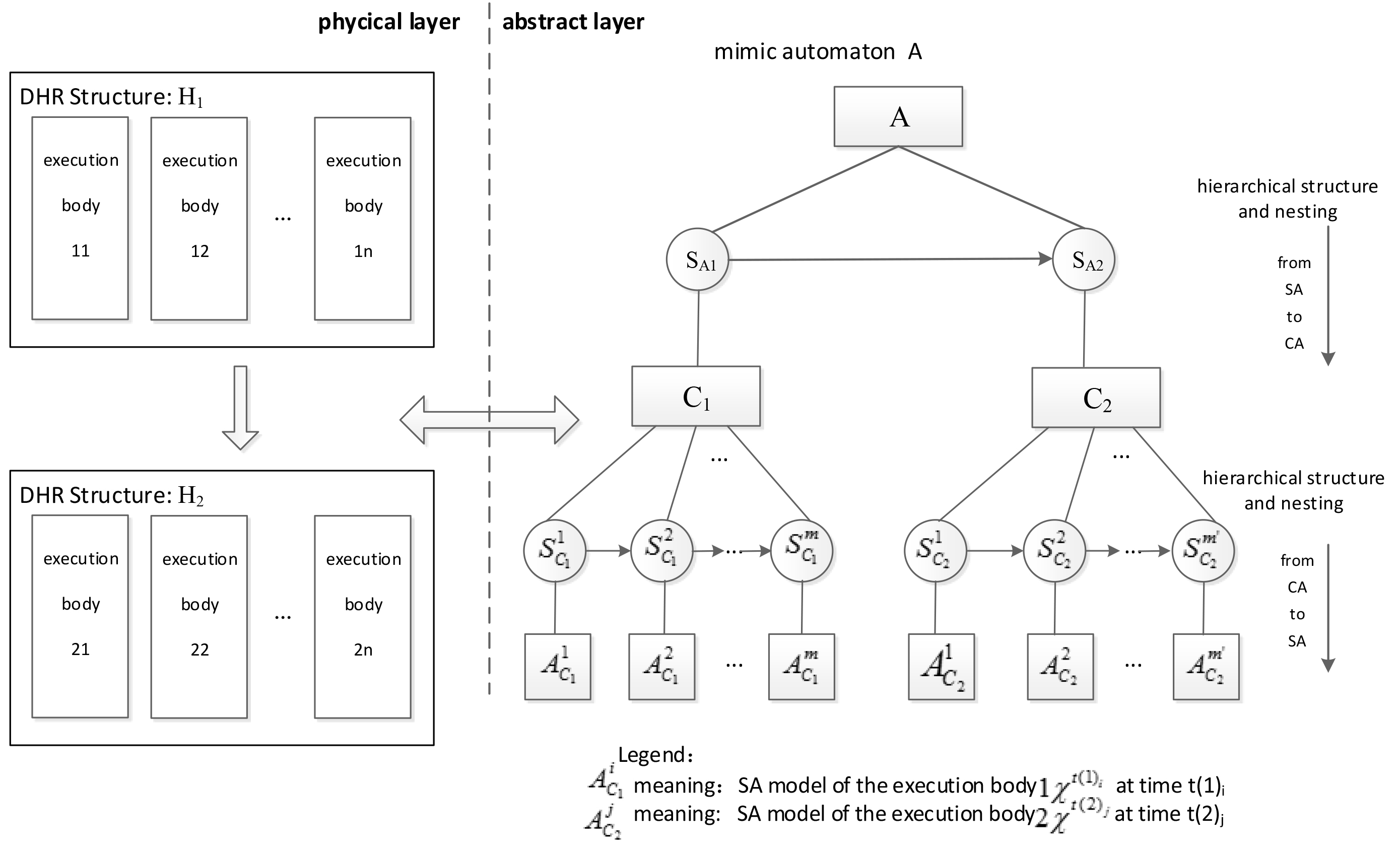}
	\caption{A mimic automaton for modeling a complex DHR structure}
	\label{fig2}
\end{figure}

A mimic automaton A is employed to model a complex DHR structure.

The left side of Fig. \ref{fig2} shows an example of a complex DHR structure. Compared with the left side of Fig. \ref{fig1}, it is composed of two DHR structures in a sequential way.

As shown in the left side of Fig. \ref{fig2}, the whole process of the mimic computing can be described as follows. At first, the computing is implemented on the first DHR structure. And then it is implemented on the second DHR structure. In other words, the first DHR structure computes for a given input. And the result of the computing, i.e., the output of the first DHR structure, is also the input for the second DHR structure. And then, the second DHR structure will compute. The result of the computing, i.e., the output of the second DHR structure, is also the result of the whole mimic computing.

We use the first DHR structure as an example. Its execution bodies are constantly changing in the whole process of computing. We assume that the first DHR structure is varied as follows. At first, the execution body 11 computes. And then, the execution body 13 computes. In this case, the input of the first DHR structure is input into the execution body 11 for computing. And the computing result of the execution body 11 is input into the execution body 13 for computing. As a result, the computing result of the execution body 13 is the output of the first DHR structure.

This is the workflow and the principle of the DHR structure. Accordingly, we can employee a mimic automaton which is shown in the right side of Fig. \ref{fig2} to establish the model for a complex DHR structure. The detailed process is portrayed as follows.

First, a cellular automaton is applied to model the dynamic changes in the structure of execution bodies. 

As illustrated in Fig. \ref{fig2}, one cellular automaton depicts one of the two DHR structures and its process of dynamic change in execution bodies, as described in example 1. As shown in the right side of Fig. \ref{fig2}, we employee the cellular automata $C_1$ and $C_2$ to describe the two DHR structures, $H_1$ and $H_2$, respectively. That is to say, $C_1$ describes the change process of the execution bodies in the DHR structure $H_1$. And $C_2$ describes the change process of the execution bodies in the DHR structure $H_2$. We use the DHR structure $H_1$ as an example. Supposing that its execution bodies change m times, $C_1$ has m states, which are $S_{C_1}^1,S_{C_1}^2,...,S_{C_1}^m$, respectively.

During the execution bodies remain constant, the ${\chi^{t(1)_i}}^{th}$ execution body of $H_1$ is responsible for computing when $C_1$ stays in the state $S_{C_1}^i$ , i.e., the period of time $(t(1)_{i-1},t(1)_i]$, where $(t(1)_{i-1},t(1)_i]$ means the time interval from the ${i-1}^{th}$ variation of execution bodies to the $i^{th}$ variation of execution bodies. Similarly, the ${\chi^{t(2)_j}}^{th}$ execution body of $H_2$ is responsible for computing when $C_2$ stays in the state  $S_{C_2}^j$, i.e., the period of time $(t(2)_{j-1},t(2)_j]$, where $(t(2)_{j-1},t(2)_j]$ means the time interval from the ${j-1}^{th}$ variation of execution bodies to the $j^{th}$ variation of execution bodies.

It is obvious that the cellular automata are not responsible for describing the specific state transitions in the computing process. And they are only responsible for describing the dynamic changes in the DHR structures.

Second, the sequential automata located in the lower layer are employed to model the specific computing process during the execution bodies remain constant.

We observe the time interval from the $2^{th}$ variation of execution bodies to the $3^{th}$ variation of execution bodies in $H_1$. At this moment, $C_1$ stays in the state $S_{C_1}^3$, as shown in the left side of Fig. \ref{fig2}. Supposing that the execution body 12 is responsible for computing, i.e., $\chi^{t(1)_3}=2$, the computing process of this execution body can be modeled by a sequential automaton $A_{C_1}^3$ in the lower layer, as shown in the right side of Fig. \ref{fig2}.

Furthermore, we observe the time interval from the $5^{th}$ variation of execution bodies to the $6^{th}$ variation of execution bodies in $H_2$. At this moment, $C_2$ stays in the state $S_{C_2}^6$, as shown in the left side of Fig. \ref{fig2}. Supposing that the execution body 28 is responsible for computing, i.e., $\chi^{t(2)_6}=8$, the computing process of this execution body can be modeled by a sequential automaton $A_{C_2}^6$ in the lower layer, as shown in the right side of Fig. \ref{fig2}.

It is obvious that the sequential automata are used to describe the specific computing process and the state transitions during the execution bodies remain constant, at the micro level.

Third, the sequential automata located in the upper layer are employed to model the logical relationship among DHR structures.

As illustrated in the left side of Fig. \ref{fig2}, the complex DHR structure is composed of the two simple DHR structures in a sequence, which is described by a sequential automaton $A_s$, where $A_s$ includes the states $S_{A1}$ and $S_{A2}$, as shown in the right side of Fig. \ref{fig2}. In other words, the sequential automaton $A_s$ in the upper layer always stays in the state $S_{A1}$ during the whole computing process of the DHR structure $H_1$. Similarly, the sequential automaton $A_s$ in the upper layer always stays in the state $S_{A2}$ during the whole computing process of the DHR structure $H_2$.

It is obvious that the sequential automata located in the upper layer only describe the logical relationship among DHR structures at the macro level.

Finally, a hierarchical automaton provides the whole skeleton for a mimic automaton. In addition, the cellular automata and the sequential automata are attached to different positions of the hierarchical automata realizing the computing of the different granularity, as shown in the right side of Fig. \ref{fig2}.
\begin{figure}
	\includegraphics[width= \textwidth]{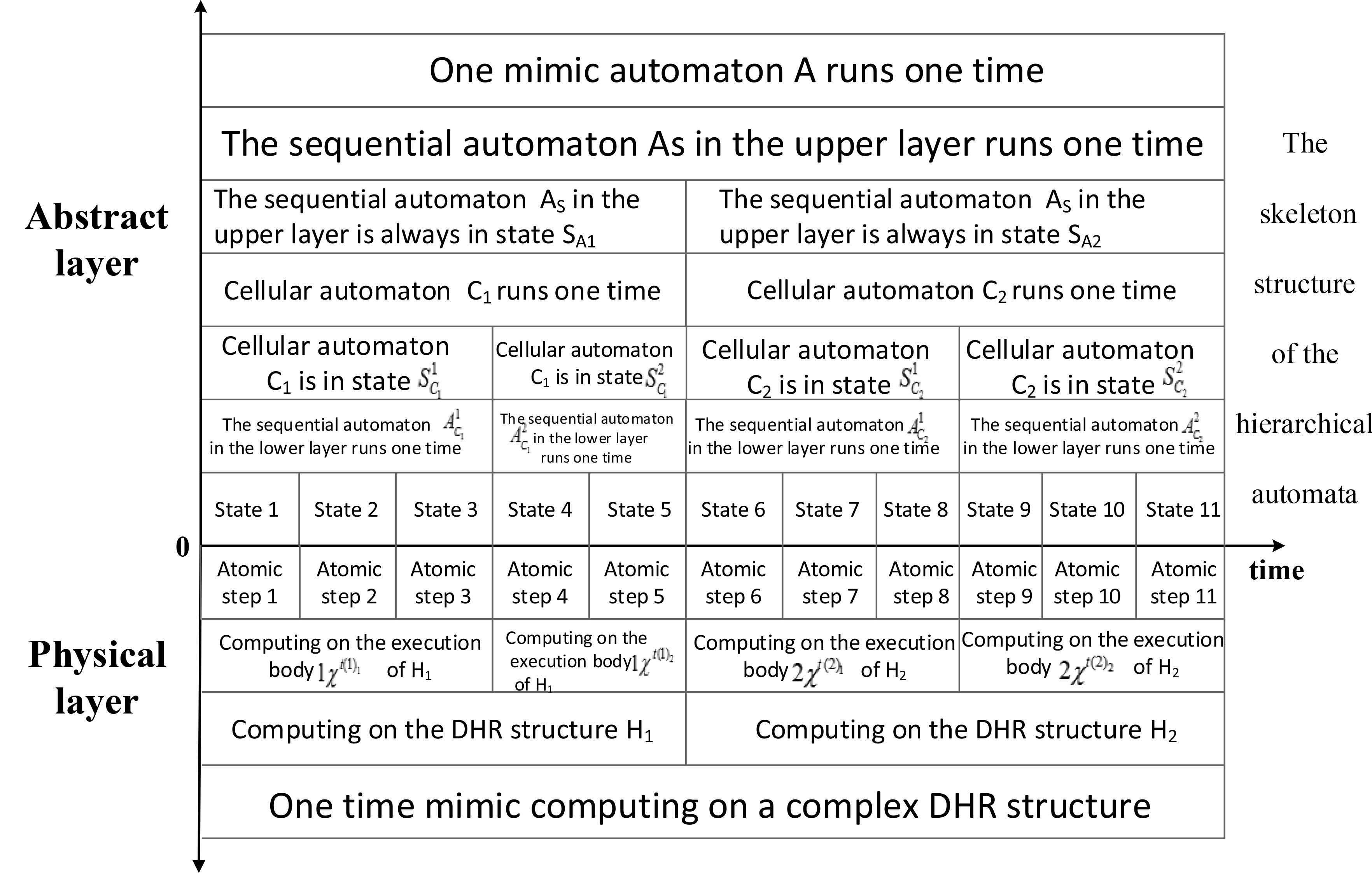}
	\caption{The relationship of time among the automata and the complex DHR structure in Fig. \ref{fig2}}
	\label{fig3}
\end{figure}

Fig. \ref{fig3} illustrates the relationship between the time interval of running of the various types of automata including the mimic automaton and the computational time interval of the various computing units.
\begin{figure}[htb!]
\centering
\subfigure[Duration 1: computing on execution body 1 at $H_1$]{
\begin{minipage}[b]{0.45\textwidth}
\includegraphics[width=\textwidth]{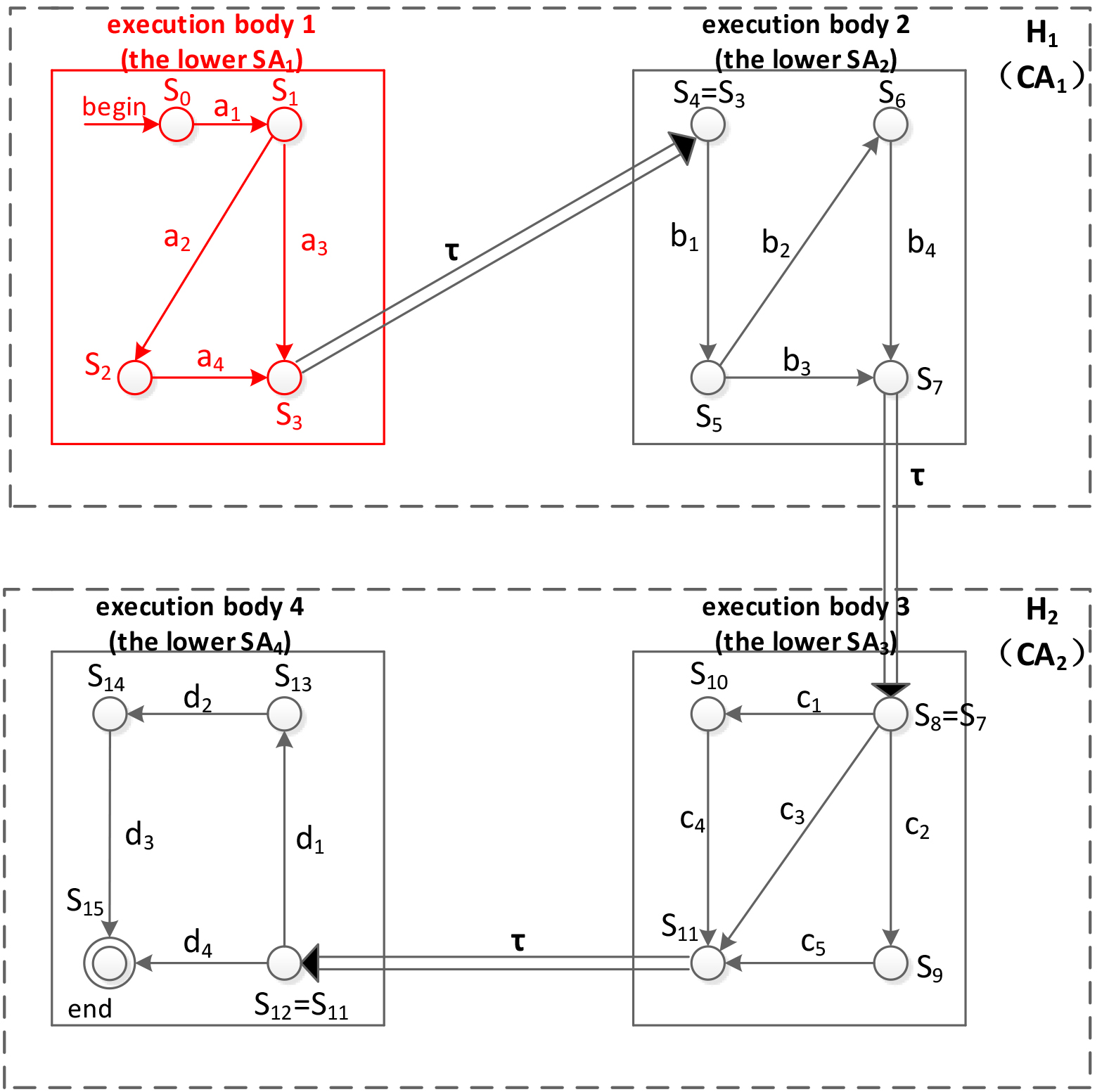} \\
\end{minipage}
}
\subfigure[ Duration 2: computing on execution body 2 at $H_1$]{
\begin{minipage}[b]{0.45\textwidth}
\includegraphics[width=\textwidth]{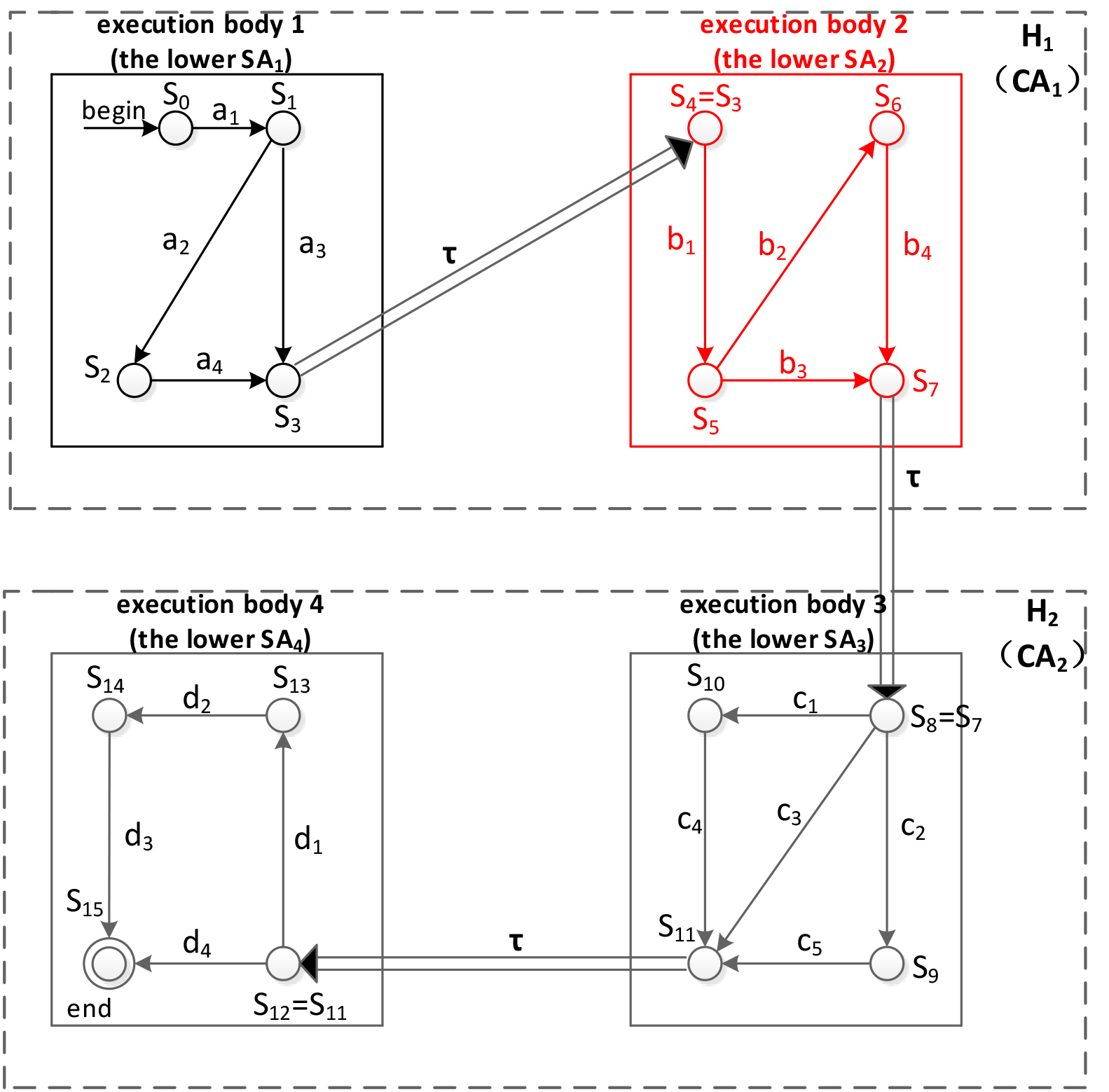} \\
\end{minipage}
}
\subfigure[ Duration 3: computing on execution body 3 at $H_2$]{
\begin{minipage}[b]{0.45\textwidth}
\includegraphics[width=\textwidth]{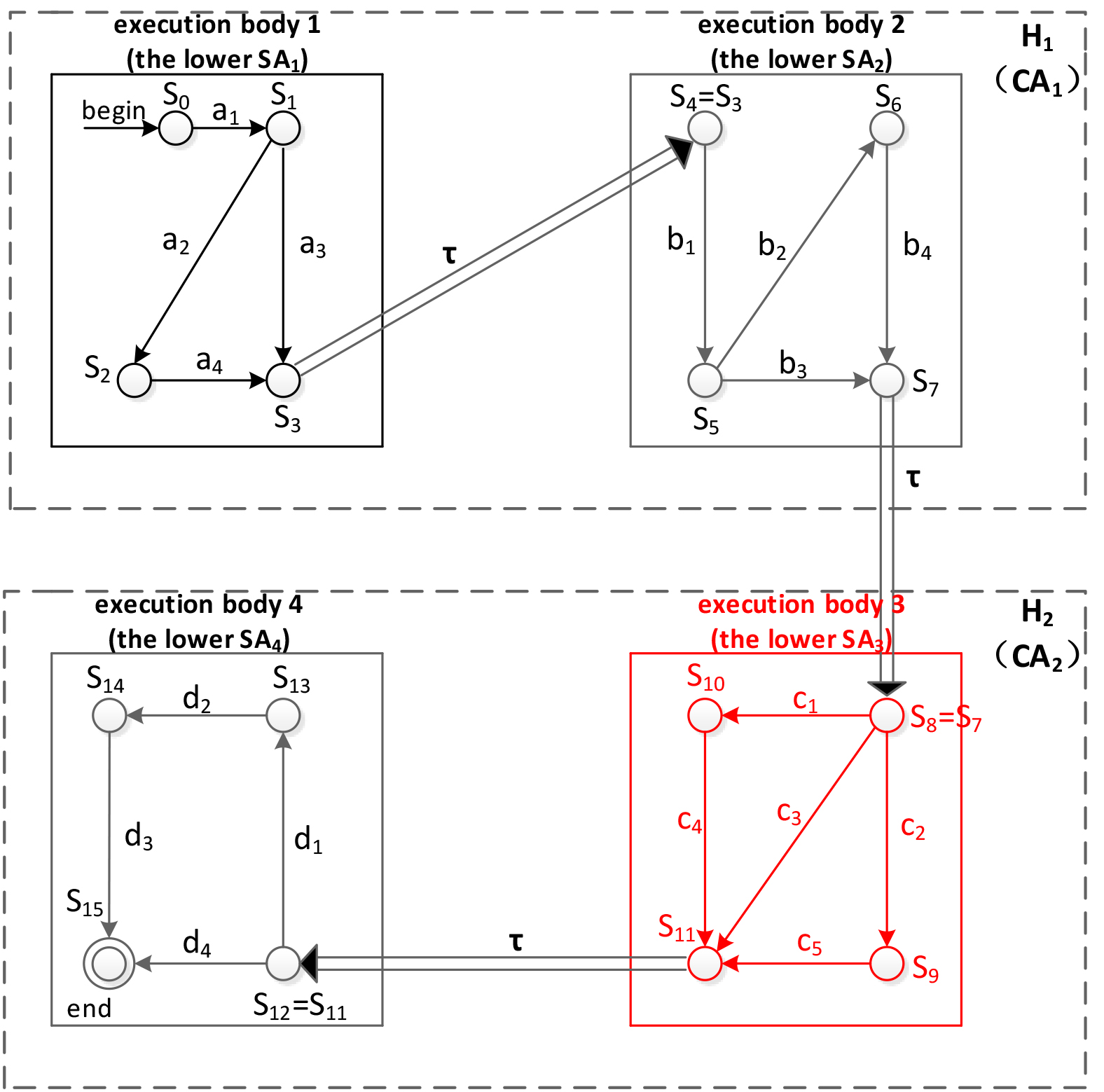} \\
\end{minipage}
}
\subfigure[ Duration 4: computing on execution body 4 at $H_2$]{
\begin{minipage}[b]{0.45\textwidth}
\includegraphics[width=\textwidth]{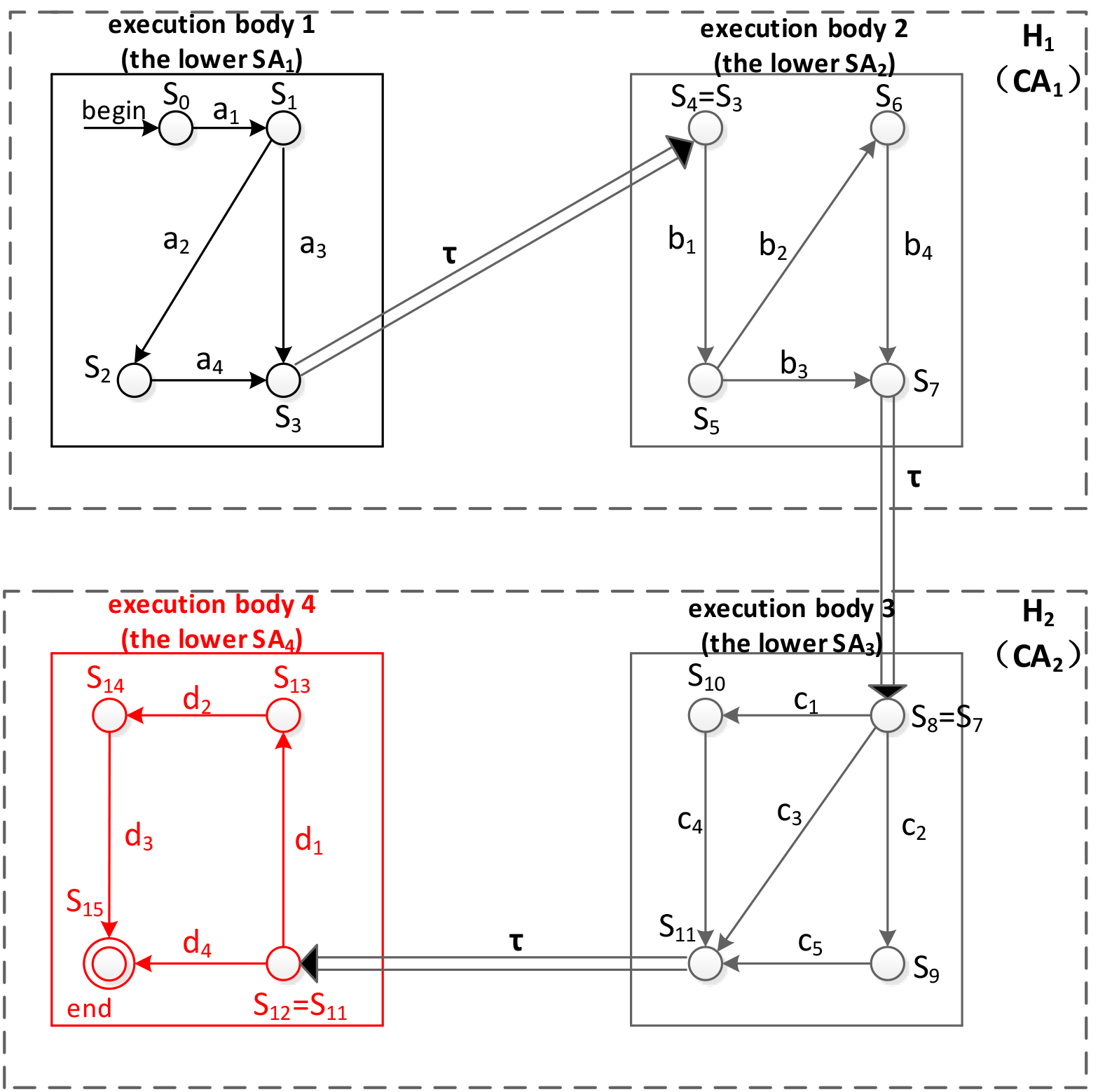} \\
\end{minipage}
}
\caption{an example: MA operation semantics (three kinds of rules of state transitions) }
\label{fig4}
\end{figure}

In this example, the HA describe the different granularity, the CA describe the dynamic reconstruction of the execution bodies at a given granularity, and the SA describe the state transitions under the circumstance of the given granularity and the given execution body. A mimic automaton which is composed of the above three types of automata, establishes a formal model for a complex DHR structure in this way. The SA in upper layer only describes a simple state transition. Thus, this SA is a finite state automaton. As for the SA in the lower layer in this example, is it a Turing machine, a linear bounded automaton, a push-down automaton, or a finite state automaton? The answer is determined by the target task to be performed in this example. In section 4, we will discuss the relationship between the target task and the selection of the various types of automata.

Fig. \ref{fig2} shows a side view of the mimic automaton in this example, in which time goes from left to right. In comparison, Fig. \ref{fig4} shows an overhead view of this mimic automaton, in which the arrows represent time and the squares represent space.

In Fig. \ref{fig4}, we illustrate the components of the mimic automaton. On the basis of it, we can analyze the operational semantics of the mimic automaton.

In Fig. \ref{fig4}, there are four SA located in the lower layer: $S_{A1}, S_{A2}, S_{A3}, S_{A4}$.

$S_{A1}$ has a state set $\{s_0,s_1,s_2,s_3\}$, a set of input alphabets $\{a_1,a_2,a_3,a_4\}$, a set of state transitions $\{s_0 \xrightarrow{a_1}s_1,s_1 \xrightarrow{a_2}s_2,s_1 \xrightarrow{a_3}s_3,s_2 \xrightarrow{a_4}s_3\}$, an initial state $s_0$, and a final state $s_3$.

$S_{A2}$ has a state set $\{s_4,s_5,s_6,s_7\}$, a set of input alphabets $\{b_1,b_2,b_3,b_4\}$, a set of state transitions $\{ s_4 \xrightarrow{b_1}s_5,s_5 \xrightarrow{b_2}s_6,s_5 \xrightarrow{b_3}s_7,s_6 \xrightarrow{b_4}s_7 \}$, an initial state $s_4$, and a final state $s_7$.

$S_{A3}$ has a state set $\{s_8,s_9,s_{10},s_{11}\}$, a set of input alphabets $\{c_1,c_2,c_3,c_4,c_5\}$, a set of state transitions $\{ s_8 \xrightarrow{c_1}s_{10},s_8 \xrightarrow{c_2}s_9,s_8 \xrightarrow{c_3}s_{11},s_{10} \xrightarrow{c_4}s_{11},s_9 \xrightarrow{c_5}s_{11}\}$, an initial state $s_8$, and a final state $s_{11}$.

Similarly, $S_{A4}$ has a state set $\{s_{12},s_{13},s_{14},s_{15}\}$, a set of input alphabets $\{d_1,d_2,d_3,d_4\}$, a set of state transitions $\{ s_{12} \xrightarrow{d_1}s_{13},s_{13} \xrightarrow{d_2}s_{14},s_{14} \xrightarrow{d_3}s_{15},s_{12} \xrightarrow{d_4}s_{15}\}$, an initial state $s_{12}$, and a final state $s_{15}$.

In Fig. \ref{fig4}, there are two cellular automata $C_{A1}$ and $C_{A2}$. For the sake of simplicity, these two cell automata are combined and denoted as CA.

CA has a state set of the four elements, i.e., $\{(1,0,0,0),(0,1,0,0),(0,0,1,0),(0,0,0,1)\}$, and a set of state transitions of  the three elements, i.e., $\{(1,0,0,0) \xrightarrow{\vert T1 \vert + \tau},(0,1,0,0),(0,1,0,0)\xrightarrow{\vert T2 \vert + \tau}(0,0,1,0),(0,0,\\1,0)\xrightarrow{\vert T3 \vert + \tau}(0,0,0,1)\}$ , where $\tau$ represents the next time point, $\vert T1 \vert$ represents the length of the time interval $T1$, i.e., the computational time on the first execution body. Furthermore, $(1,0,0,0)$ means that CA is staying the following state: the first execution body is computing.

On the basis of it, we can define the MA operational semantics, which has the three sets of state transition rules.
\begin{enumerate}[(1)]
 \item The first set of rules is related to the internal state transitions of the SA in lower layer, including the following rules: $<a_1,s_0>\xrightarrow {} s_1, <a_2,s_1>\xrightarrow {} s_2,<a_3,s_1>\xrightarrow {} s_3,<a_4,s_2>\xrightarrow {} s_3,<b_1,s_4>\xrightarrow {} s_5,<b_2,s_5>\xrightarrow {} s_6,<b_3,s_5>\xrightarrow {} s_7,<b_4,s_6>\xrightarrow {} s_7,<c_1,s_8>\xrightarrow {} s_{10},<c_2,s_8>\xrightarrow {} s_9,<c_3,s_8>\xrightarrow {} s_{11},<c_4,s_{10}>\xrightarrow {} s_{11},<c_5,s_9>\xrightarrow {} s_{11},<d_1,s_{12}>\xrightarrow {} s_{13},<d_2,s_{13}>\xrightarrow {} s_{14},<d_3,s_{14}>\xrightarrow {} s_{15},<d_4,s_{12}>\xrightarrow {} s_{15}$. We take $<a_1,s_0>\xrightarrow {} s_1$  as an example, it indicates that an input $a_1$, which is an atomic step of program, transfer the state from $s_0$ to $s_1$.
 \item The second set of rules is related to the internal state transitions of CA, including the following rules: $<(\vert T1 \vert +\tau),(1,0,0,0)>\xrightarrow {}(0,1,0,0), <(\vert T2 \vert +\tau),(0,1,0,0)>\xrightarrow {}(0,0,1,0),<(\vert T3 \vert +\tau),(0,0,1,0)>\xrightarrow {}(0,0,0,1)$. We take $<(\vert T1 \vert +\tau),(1,0,0,0)>\xrightarrow {}(0,1,0,0)$ as an example, it represents that the state is transformed from $(1,0,0,0)$ to $(0,1,0,0)$, after time lapses $\vert T1 \vert +\tau$ unit, which is an atomic step of CA.
 \item The third set of rules is related to the state transitions among the SA in lower layer, including the following rules: $<\tau ,s_3> \xrightarrow {} s_4=s_3,<\tau ,s_7> \xrightarrow {} s_8=s_7,<\tau ,s_{11}> \xrightarrow {} s_{12}=s_{11}$.We take $<\tau ,s_3> \xrightarrow {} s_4=s_3$ as an example, it means that the state is transformed from $s_3$ to $s_4=s_3$, after time lapses $\tau$ unit, which is an empty atomic step of SA.  
\end{enumerate}

The transition rules at a higher level of abstraction, such as the internal state transitions of SA in the upper layer and the state transitions among CA, are implementation-agnostic of the machine. Thus, their operational semantics do not need to be defined. In fact, the transition rules at a upper level of abstraction can be implemented by calling the above three sets of rules of operational semantics. Therefore, these three sets of rules can define the semantics of the running of MA. $\Box$
\section{The computational ability of mimic automata}
The mimic automata for modeling the different mimic computing systems have the different computational abilities. And a mimic automaton for modeling a general mimic computer is equivalent intuitively to the Turing machine. 

{\bf Proposition 1} The complete version of a mimic automaton is equivalent to the Turing machine.

The key idea of proof is as follows.

$\bot$: According to the definition of MA in section 2, the Turing machine is a component of the complete version of a mimic automaton. Thus, the lower bound of the computational ability of the complete version of the mimic automaton is a Turing machine.

$\top$:

\begin{enumerate}[]
\item[\textcircled{1}] The existing studies have revealed that the formal language accepted by the four types of component automata of MA does not exceed the scope of the recursively-enumerable language.
\item[\textcircled{2}]  According to the definition of MA in section 2, the compositional way of component automata does not exceed the scope of recursively enumerable language.
\item[\textcircled{3}] According to \textcircled{1} and \textcircled{2}, the formal language accepted by MA does not exceed the scope of the recursively enumerable language. Therefore, the upper bound of the computational ability of MA is a Turing machine.
\end{enumerate}

According to the proof of $\bot$ and $\top$, both the upper bound and the lower one of the computational ability of MA is a Turing machine. Therefore, the complete version of MA is equivalent to a Turing machine in terms of the computational ability, according to the Sandwich Theorem. $\Box$

{\bf Corollary 2} The formal language accepted by a complete version of mimic automaton is a recursively enumerable language.

In fact, the different situations on the power of MA can be discussed as follows, if a mimic automaton and its engineering model of the mimic computing are associated.

Case 1: A MA is a standard Turing machine, if one of SA is a Turing machine, and the CA are empty. The engineering model described by this formal model is a traditional computer. Under this circumstance, the mimic computing architecture does not make any dynamic change. And the mimic computing degenerates into the traditional computing.

Case 2: If the SA is a LBA, and the CA is a standard one-dimensional cellular automaton, the engineering model described by the MA formal model may be the following one. A program runs on an execution body, and the variant relationship among the execution bodies presents in the logical way of the recursively enumerable language. Under this circumstance, the dynamic change of the mimic computing architecture is very complicated, which is equivalent to the Turing machine. On the one hand, the hardware structure is dynamic changed in a complex way. On the other hand, the program is performed in a traditional way during the hardware structure remains constant.

Case 3: If the SA is a LBA, the CA is limited to describe the computing phenomenon which is specified by a FSA, the engineering model described by the MA formal model may be the following one. A program runs on an execution body, and the variant relationship among the execution bodies satisfies the property of Regular Expression (RE). Under this circumstance, the mimic computing architecture varies dynamically and coincides with the RE property. On the one hand, the hardware structure is dynamic changed in a simple way. On the other hand, the program is performed in a traditional way during the hardware structure remains constant.

Case 4: If the SA is a FSA, the CA is limited to describe the computing phenomenon which is specified by a FSA, the engineering model described by the MA formal model may be the following one. A program written by a weak language runs on an execution body, and the variant relationship among the execution bodies satisfies the RE property. Under this circumstance, the mimic computing architecture varies dynamically and coincides with the RE property. On the one hand, the hardware structure is dynamic changed in a simple way. On the other hand, the program written by a weak language is performed in a traditional way during the hardware structure remains constant.
\section{Conclusions}
In this paper, a mimic automaton is constructed to establish a formal model for the mimic computing phenomenon in a typical application scenario. And the operational semantics and the computational ability of MA are analyzed. As a result, some conclusions about the MA theory are obtained. This is the main contribution of this paper.

\section*{Acknowledgements}

This work has been supported by the National Natural Science Foundation of China (No.U1204608).

\end{document}